\documentclass[prd,twocolumn,notitlepage,longbibliography,nofootinbib,superscriptaddress,preprintnumbers]{revtex4-1}
\usepackage[utf8]{inputenc}

\usepackage{bm}
\usepackage{comment} 
\usepackage[colorlinks=true,urlcolor=blue,anchorcolor=blue,citecolor=blue,linkcolor=blue,pagecolor=blue]{hyperref}
\usepackage{amsmath,latexsym,amssymb,mathrsfs}
\usepackage{graphicx}

\begin{document}
\title{Global Spacetime Structure of Compactified Inflationary Universe}
\author{Tokiro Numasawa}
\email{tokiro.numasawa@physics.mcgill.ca}
\affiliation{Department of Physics, McGill University, Montr\'eal, Qu\'{e}bec, H3A 2T8, Canada}
\affiliation{Department of Physics, Graduate School of Science,
Osaka university, Toyonaka 560-0043, Japan}
\author{Daisuke Yoshida}
\email{d.yoshida@physics.mcgill.ca}
\affiliation{Department of Physics, McGill University, Montr\'eal, Qu\'{e}bec, H3A 2T8, Canada}
\preprint{OU-HET 991}
\begin{abstract}
We investigate the global spacetime structure of torus de Sitter universe, which is exact de Sitter space with torus identification based on the flat chart.
We show that past incomplete null geodesics in torus de Sitter universe are locally extendible. Then we give an extension of torus de Sitter universe so that at least one of the past incomplete null geodesics in the original spacetime becomes complete. However we find that extended torus de Sitter universe has two ill behaviors. The first one is a closed causal curve. The second one is so called quasi regular singularity, which means that there is no global, consistent extension of spacetime where all curves become complete, nevertheless each curve is locally extensible.
\end{abstract}
\maketitle

\section{Introduction}
Is our Universe finite or infinite?
Since our Universe is approximately homogeneous and isotropic, it should be effectively described by Friedmann-Lema\^{i}tre-Robertson-Walker (FLRW) spacetime. The global structure of FLRW spacetime can be completely determined by the sign of its spatial curvature, provided that it is simply connected. If the spatial curvature is positive (closed FLRW spacetime), each spatial slice is 3 dimensional sphere $S^3$, which has a finite volume. On the other hand if spatial curvature is zero (flat FLRW spacetime) or negative (open FLRW spacetime), each spatial slice is isometric to Euclid space $\mathbb{R}^3$ or hyperbolic space $H^3$ respectively, both of which have an infinite volume. These are, however, not the all cases.
Interestingly, we can also construct flat or open universe with finite volume by assuming non-trivial topology of universe.
 Possibility of such compact topology has not been excluded any observational and theoretical considerations. Rather, a compact universe is favored, e.g., in a scenario of quantum creation of a universe~\cite{Zeldovich:1984vk, Coule:1999wg, Linde:2004nz} because there is no potential barrier in the Wheeler DeWitt equation for the flat and open cases.

In the present paper, we focus on flat FLRW spacetime,
\begin{align}
 g_{\mu\nu}dx^{\mu}dx^{\nu} = -d t^2 + a(t)^2 \delta_{ij} dx^{i} dx^{j},\label{FLRW}
\end{align}
where $t$ is the proper time of comoving observers, $x^i$ $(i = 1,2,3)$ are the comoving coordinates and $a(t)$ is the scale factor.   
Since flat FLRW Universe is invariant under the translation $x^{i} \rightarrow x^{i} + R^{i}$ with constants $R^i$, there is no contradiction with the identification, 
\begin{align}
 x^i \sim x^i + R^i.\label{torusid}
\end{align}
Under this identification, a $t =$ constant slice describes compact space, which is 3 dimensional torus $T^3$ with periods $R^i$. This spacetime is called ``three-torus universe''~\cite{Zeldovich:1984vk}.  According to the recent results of Planck~\cite{Ade:2015bva}, a lower bound of the periods of the three-torus universe is given by $R^{i}/2  > 0.97 \chi_{rec}$, where $\chi_{rec}$ is the distance to the recombination surface. The purpose of the present paper is to discuss the singularity problem for the three-torus universe in an inflationary epoch.

Although no singularity exists in exact de Sitter space, the absence of singularities in inflationary universe is still non-trivial even without the torus identification. Actually it was pointed out that inflationary universe has past incomplete geodesics by Ref.~\cite{Borde:2001nh}. This result is reasonable because an inflationary universe is approximately described by {\it flat} de Sitter space, not entire de Sitter space, and flat de Sitter space contains a past incomplete geodesic, which is just a part of the corresponding past complete geodesic in entire de Sitter space. We stress that, as an example of the flat de Sitter case, the presence of a past incomplete curve itself does not mean spacetime singularities. It corresponds to the presence of the past boundary, which can be reached along a geodesic with a finite affine parameter. Then, a spacetime is said to have a singularity if a boundary is inextendible~\cite{Hawking:1973uf, Ellis:1977pj} \footnote{
In other words, an {\it inextendible} spacetime is said to have singularity if it contains an incomplete geodesic.}.

The main goal of this paper is to clarify the extendibility of inflationary torus universe beyond the past boundary. When we consider the model of inflationary torus universe, there are two big differences from exact de Sitter space that we need to care about: First, the scale factor of a real inflationary universe is not exactly de Sitter one, otherwise inflation can not end. Second, there is the torus identification, which is the main topic of this paper. 

The extensibility of the past boundary due to the difference of scale factors was recently clarified by Ref. \cite{Yoshida:2018ndv}. There it was shown that the past boundary becomes so-called parallely propagated curvature singularity if $\dot{H}/a^2$ diverges as $t \rightarrow - \infty$, where $\dot{H}$ is the time derivative of the Hubble function. On the other hand if it converges, an inflationary universe can be continuously extended at least locally. We refer to \cite{Hawking:1973uf, Ellis:1977pj} for the precise definition of parallely propagated curvature singularity and local extendibility. 

In the present paper, then, we will focus on how the extensibility beyond the past boundary is affected by the torus identification \eqref{torusid}.  To focus on this question, we ignore the deviation of the metric from the exact de sitter one.
Therefore, the scale factor is given by
\begin{align}
 a(t) = \mathrm{e}^{H t},
\end{align}
rather than that of a realistic inflationary universe. We would like to emphasize that, although it was known that torus de Sitter universe contains incomplete geodesics (See, e.g. ,Refs.~\cite{Galloway:2004bk, Galloway:2017mts}), extendibility of these incomplete curves has not been discussed in any literatures. 

Our paper is organized as follows.
In the next section we will demonstrate the past incompleteness of a torus de Sitter universe. Then in the section \ref{sec:extension} we will present a way to extend a torus de Sitter universe beyond its past boundary. After that, we will however find two ill behaviors of the extended torus de Sitter universe: a closed null curve and quasi regular singularities. Final section is devoted to a summary and discussion.  
For later convenience, we summarize how to call each spacetime in the table~\ref{hyou1}.
\begin{table}[htbp]
\begin{tabular}[t]{c|cc}
 & before extension& after extension\\
\hline
uncompactified
&flat de Sitter & (entire/closed) de Sitter \\
compactified &torus de Sitter & extended torus de Sitter \\
\end{tabular}
\caption{summary of how to call each spacetime}
\label{hyou1}
\end{table}
         
\section{Past incomplete geodesics in torus de Sitter universe}
\begin{figure}[htbp]
\begin{center}
\includegraphics[width=\hsize]{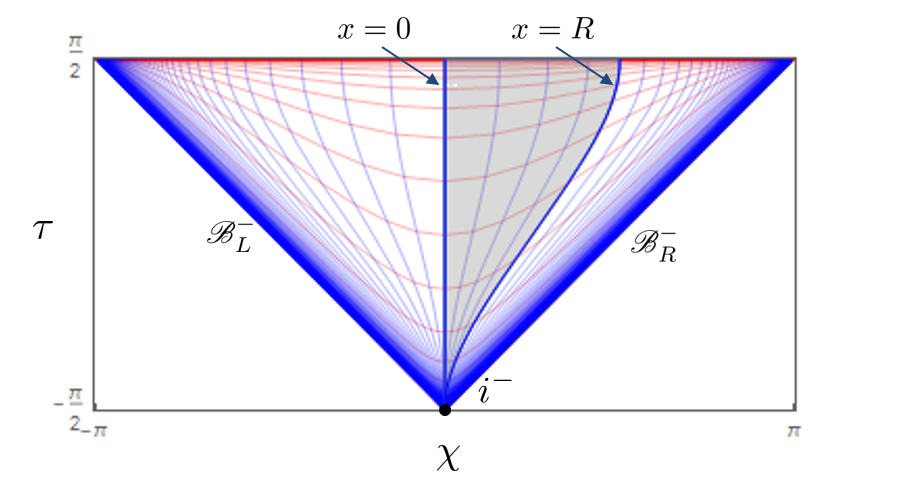}
\caption{Conformal diagram of flat and torus de Sitter spacetime
 on the $x^2 = x^3 = 0$ surface:
The red and blue curves represent $t$ and $x$ constant curves respectively. The flat chart only covers upper half triangle region of the entire de Sitter spacetime. The gray colored region represents a fundamental region of torus de Sitter universe.
 }
\label{dsf}
\end{center}
\end{figure}

In this section, we demonstrate the presence of incomplete geodesics in torus de Sitter universe.  
For our purpose, it is sufficient to consider the 2 dimensional spacetime region $x^2 = x^3 = 0$ in a flat de Sitter space. Then the metric on this region is given by
\begin{align}
 ds^2|_{x^2=x^3=0} = -dt^2 + \mathrm{e}^{2 H t} (dx)^2,
\end{align}
where $x$ represents $x^1$.
The global spacetime structure can be seen by taking further coordinate transformation to the coordinates $\{ \tau, \chi \}$, which are given as
\begin{align}
 &\cos \chi = \zeta(t,x) \left( 1 + \mathrm{e}^{2 H t}(1 - H^2 x^2) \right) ,\notag\\
 & \sin \chi = \zeta(t,x)~  2 H \mathrm{e}^{2 H t} x ,\notag\\
 &\tan \tau = \frac{- 1 + \mathrm{e}^{2 H t} (1 + H^2 x^2)}{2 \mathrm{e}^{H t}}, \label{eq:ds2coordchange}
\end{align}
where
we introduced $\zeta(t,x)$ simply to avoid complicated expressions. The concrete expression of $\zeta(t,x)$ can be obtained from the relation $\cos^2\chi + \sin^2 \chi = 1$.
Here we solve these equations in the region $\chi \in (-\pi, \pi)$ and $\tau \in (-\pi/2 , \pi/2)$.
In these coordinates, the metric becomes conformally flat form,
\begin{align}
 ds^2|_{x^2=x^3=0} = \frac{-d\tau^2 + d \chi^2}{H^2 \cos^2 \tau}.
\end{align}
As well known, and as shown explicitly in appendix A,
entire de Sitter space corresponds to the region $\chi \in (-\pi, \pi)$ and $\tau \in (-\pi/2 , \pi/2)$ and flat de Sitter space corresponds to a half of it. The conformal diagram is written in Fig.\ref{dsf}. 
In the 2 dimensional region, the past boundary is separated into two disconnected boundaries ($\tau = \pm \chi - \pi/2$), which we call $\mathscr{B}_R^-$ and $\mathscr{B}_L^-$.

Now we consider the torus identification along $x$ direction,
\begin{align}
 x \sim x + R,  \qquad \text{(along $t = $ const.)}. \label{S1}
\end{align}  
Then conformal diagram is given just by identifying the points related through \eqref{S1} in Fig.\ref{dsf}. The original flat de Sitter space can be regarded as infinite copies of an fundamental region, e.g. $x \in [0, R) , t \in (-\infty, \infty)$ which is the gray colored region in Fig. \ref{dsf}. Note that the conformal diagram of torus FLRW Universe can be found in Ref.~\cite{Ellis:2015wdi}.  

To examine the incompleteness of torus de Sitter spacetime,
we focus on the curve $\gamma_R(\lambda)$ which defined by
\begin{align}
 t(\lambda) = \frac{1}{H} \log (H \lambda), \qquad
 x(\lambda) = - \frac{1}{H^2 \lambda},
\end{align}
with $x^2=x^3=0$.
This curve propagated to $+ x$ direction as $t$ goes to $- \infty$. One can directory show that this curve is actually an affine parametrized null geodesic because the tangent vector of this curve,
\begin{align}
 k^{\mu}\partial_{\mu}  = \frac{1}{H \lambda} \partial_{t} + \frac{1}{H^2 \lambda^2} \partial_x,
 \end{align}
is null and satisfies the affine parametrized geodesic equation,
\begin{align}
 k^{\mu}\nabla_{\mu} k^{\nu} = 0.
\end{align}
Clearly this curve is past incomplete in the flat de Sitter space because its affine parameter $\lambda$ is finite in the limit $t \rightarrow -\infty$, which is the limit to $\mathscr{B}^{-}_{R}$. This is not surprising because this curve is simply a part of a complete null geodesic in the entire de Sitter space.

\begin{figure}[htbp]
\begin{center}
 \includegraphics[width=\hsize]{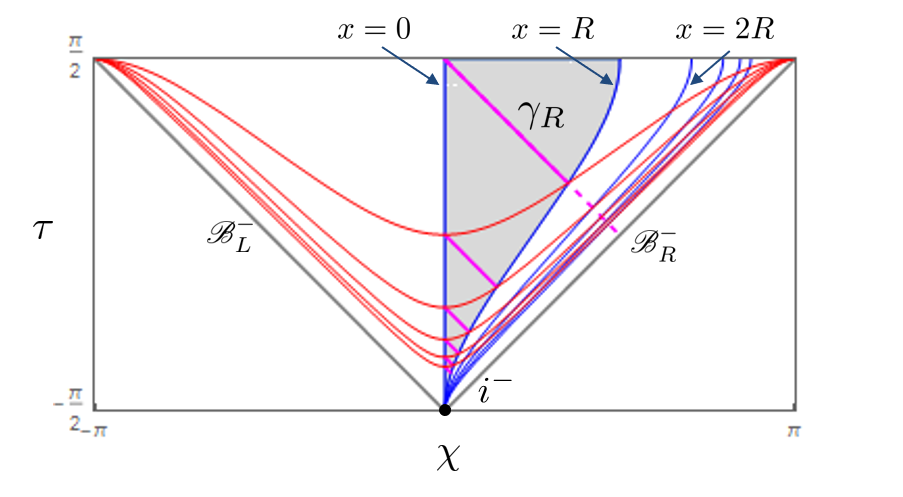}
\caption{Incomplete null geodesic in torus de Sitter Universe:
The left and right boundary of gray colored region ($x=0$ and $x = R$) are identified along the red colored curves ($t=$ constant). An incomplete null geodesic $\gamma_{R}$ in the flat de Sitter region toward $\mathscr{B}^{-}_{R}$  is represented as a null curve (solid magenta one) toward $i^-$ in a fundamental region. }
\label{dsfnull}
\end{center}
\end{figure}

These curves remain incomplete even after the torus identification \eqref{S1} is imposed, as suggested by more general analysis for spacetimes with compact Cauchy surfaces \cite{Galloway:2004bk, Galloway:2017mts}.  
In the fundamental region this incomplete null geodesic looks going to the past timelike infinity $i^{-}$ while it turns around the torus infinite times, as written in Fig.\ref{dsfnull}. 
Apparently it seems non-trivial whether the spacetime can be extended so that this curve becomes complete. In the next section we will see that (local) extendibility is clarified if we use another fundamental region of the spacetime.
 
\section{Extension of torus de Sitter Universe}
\label{sec:extension}
\begin{figure}[htbp]
\begin{center}
 \includegraphics[width=\hsize]{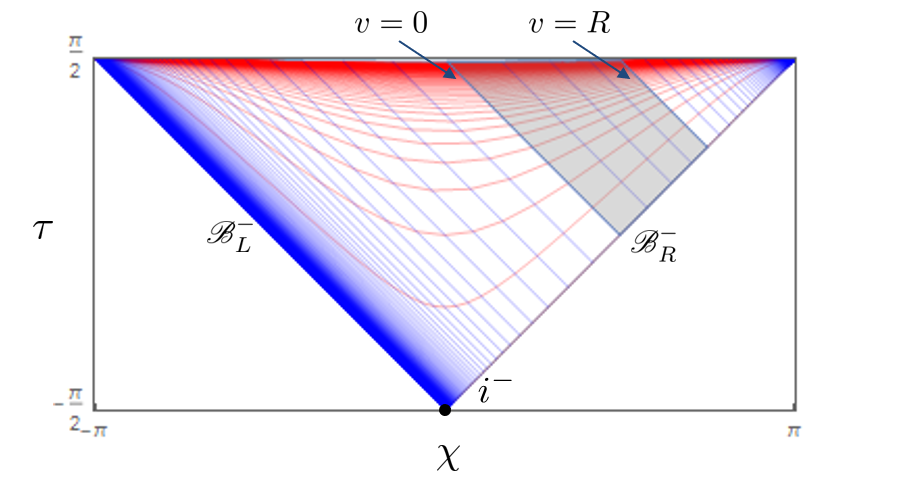}
\caption{The Penrose diagram of flat and torus de Sitter space in $(\lambda, v)$ coordinates:
The red  and blue curves represent $\lambda =$ constant and $v =$ constant surfaces respectively. One can use the gray colored region as a fundamental region of torus de Sitter Universe. 
}
\label{fdsvl}
\end{center}
\end{figure}

Since the original flat de Sitter space can be extended beyond past boundary $\mathscr{B}^{-}_R$, it might be possible to extend the flat torus de Sitter universe beyond its boundary. Let us introduce following new coordinates by an analogy to the Eddington-Finkelstein coordinates in Schwarzschild spacetime;
\begin{align}
 &\lambda = \frac{1}{H}\mathrm{e}^{H t}, \notag \\
 &v =  x - \frac{1}{H} \mathrm{e}^{-H t}.
\end{align}
Here $\lambda$ is an affine parameter of a past right going null geodesic characterized by $v =$ constant.
See Ref.\cite{Yoshida:2018ndv} for applications of these coordinates to the discussion about the extendibility beyond $\mathscr{B}^{-}_R$ for inflationary universe with trivial topology. 
In these coordinates, the metric of the 2 dimensional region of the flat de Sitter space can be expressed as
\begin{align}
 ds^2|_{x^2=x^3=0} &= - 2 dv\, d\lambda + H^2 \lambda^2 dv^2  \label{gvlam}
\end{align}
The region covered by the flat chart corresponds to the region $\lambda \in (0, \infty)$, and the Penrose diagram is given by Fig. \ref{fdsvl}.
Since $t=$ constant surface coincides with $\lambda =$ constant one, the torus identification \eqref{S1} can be expressed in terms of the new coordinates as
\begin{align}
 v \sim v + R, \qquad  \text{(along $\lambda =$ const.)}.\label{torusvlam}
\end{align}
Based on these coordinates, it is natural to use the region $v \in [0, R) , \lambda \in (0, \infty)$ as a fundamental region of torus de Sitter universe, which is the gray colored region in Fig. \ref{fdsvl}.

\begin{figure}[htbp]
\begin{center}
 \includegraphics[width=\hsize]{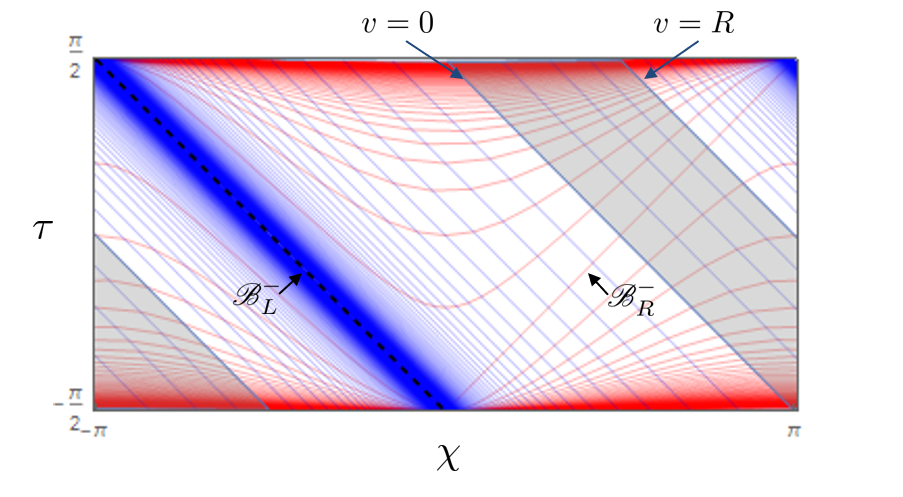}
\caption{Penrose diagram of the extended torus de Sitter universe:
Now $(\lambda,v)$ coordinates cover the region of the entire de Sitter space except for $\mathscr{B}^{-}_{L}$.
Note that the right and left ends of the diagram are identified.}
\label{dSext} 
\end{center}
\end{figure}

By the use of this fundamental region, it is clear that past right going null geodesics can be extended by the same way as the flat de Sitter space with trivial topology. 
We already have an explicit way to extend beyond the boundary $\mathscr{B}_R^-$: the coordinate $(\lambda, v)$ provides such extension, because the metric \eqref{gvlam} is regular in the region $\lambda \in (- \infty ,  \infty)$ and $v \in (-\infty, \infty)$. 
The important point here is that  the torus identification \eqref{torusvlam} is also well defined for any value of $\lambda \in (-\infty, \infty)$. 
Thus now torus de Sitter universe is extended beyond $\mathscr{B}^{-}_R$. Since $\lambda$ is nothing but the affine parameter of the past right going null geodesics, past right going null geodesics are complete there.
The Penrose diagram of the extended torus de Sitter space is given by Fig. \ref{dSext}.

\section{Property of extended torus de Sitter universe}
We gave an extension of torus de Sitter universe and confirmed a past right going null geodesic $\gamma_{R}$ are past complete in the previous section. However there still exists two ill behavior, which are closed causal curve and quasi regular singularity, as we will see below.  
\subsection{Closed causal curves}
\begin{figure}[htbp]
\begin{center}
 \includegraphics[width=\hsize]{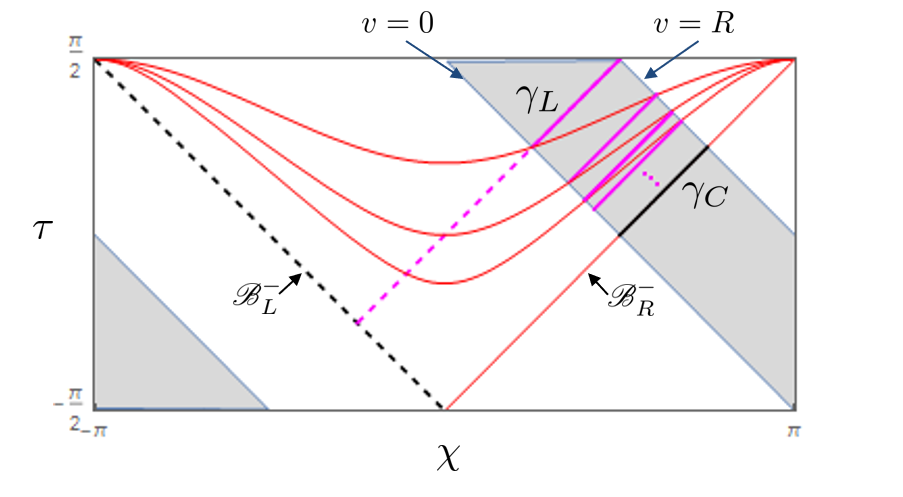}
\caption{Closed causal curve and incomplete curve:
The black solid line represents the closed null curve $\gamma_{C}$. 
The past left going incomplete null geodesic in the flat de Sitter space, which is the partly dashed magenta curve toward $\mathscr{B}^{-}_L$, is mapped to the solid magenta curve $\gamma_{L}$ in the fundamental region.
Though this curve approaches to $\mathscr{B}^{-}_R$, there is no extension where the resultant curve across $\mathscr{B}^{-}_R$.  
 }
\label{dSclnc}
\end{center}
\end{figure}

One can find that the curve $\lambda = 0$ with parameter $v$ is an affine parametrized null geodesic as it can be directly checked that its tangent vector,
\begin{align}
 l^{\mu}\partial_{\mu} := \partial_{v},
\end{align}
is null; 
\begin{align}
 g_{\mu\nu}l^{\mu} l^{\nu}|_{\lambda = 0} = g_{vv} |_{\lambda = 0} = 0,
\end{align}
and satisfies the affine parametrized geodesic equation; 
\begin{align}
 l^{\mu} \nabla_{\mu} l^{\nu}|_{\lambda=0} = 0
.
\end{align}
However our identification rule \eqref{torusvlam} implies that this curve is closed. This corresponds to the black curve $\gamma_C$ on $\mathscr{B}^{-}_{R}$  in Fig. \ref{dSclnc}. Thus causality is ill defined in our extended torus de Sitter universe. 

\subsection{Quasi regular singularities}
We presented the extension of torus de Sitter universe so that past right going null geodesics are complete. 
However it is still nontrivial whether past {\it left} going null geodesics are complete.
An incomplete left going null geodesics $\gamma_L$ is mapped to our fundamental region as in Fig. \ref{dSclnc}. As is clear from the Fig. \ref{dSclnc}, it turns around torus infinite times toward $\mathscr{B}^{-}_R$.  

Then is it possible to extend this incomplete geodesic $\gamma_{L}$? Since a point on $\mathscr{B}_R^{-}$ is an accumulation point of this curve, if this curve is extensible, it has to through a point on $\mathscr{B}_R^{-}$
\footnote{
\label{FN2}
Here we implicitly assumed that the spacetime is Hausdorff.  If one allows a non-Hausdorff spacetime, it is possible that $\gamma_L$ converges a point other than $\mathscr{B}^{-}_{R}$.
Similar structure was known in Misner spacetime \cite{Hawking:1973uf}.
}
. However one can show that this is impossible because only the null geodesics on the $x^2=x^3=0$ plane which start from the point on $\mathscr{B}_R^{-}$ are the curve $v=$ constant (past right going one) and $\lambda = 0$ (closed null curve discussed above).  Neither of them coincides with the curve $\gamma_L$. Thus there is no extension of this incomplete curve.

We would like to stress that there is no preference between the past left and right going null geodesics in torus de Sitter universe before extension. Thus we can extend a torus de Sitter universe so that past left going null geodesics, i.e. $\gamma_{L}$, are complete by the same way we have presented here, if we do not impose the completeness of past right going null geodesic $\gamma_{R}$. This means that the end points of our incomplete geodesics are locally regular. Actually, no curvature invariants, as well as no components of curvature tensor with respect to parallely propagated basis, diverge there. This kind of singularity is called quasi regular singularity \cite{Ellis:1977pj}. Similar structure of singularity is known in, for example, Taub-NUT spacetime \cite{Taub:1950ez, Newman:1963yy} and Misner spacetime \cite{1967rta1.book..160M}. 

\subsection{Comparison with Misner spacetime }
The spacetime structure of torus de Sitter Universe is very similar to that of Misner spacetime. The similarity can be more apparent if one write the metric of torus de Sitter space in the coordinates $\{\lambda, x\}$,
\begin{align}
 ds^2|_{x^2=x^3=0} = - \frac{1}{\lambda^2} d \lambda^2
 +  \lambda^2 dx^2,\label{dsds}
\end{align}
where we set $H = 1$ for simplicity. This coordinate is singular at $\lambda = 0$ and the region $\lambda \in (0, \infty)$ covers the same region as flat de Sitter space.  On the other hand, the metric of Misner spacetime is given by
\begin{align}
 ds^2_{\text{Misner}} = - \frac{1}{\lambda} d\lambda^2 + \lambda d x^2,\label{dsmisner}
\end{align}
where $\lambda \in (0, \infty)$ and $x$ is an angular coordinate on $S^1$, namely two point related by $x \sim x + 2\pi$ are identified. The universal covering of Misner spacetime is actually the inside of the Rindler future horizon in Minkowski spacetime, which is called Milne universe\footnote{The coordinates $\{\lambda, x\}$ of the Milne Universe are related with the global Minkowski coordinates $\{T,X\}$ through $T = 2 \sqrt{\lambda} \cosh x/2$ and  $X = 2 \sqrt{\lambda} \sinh x/2$.}. By comparing \eqref{dsds} and \eqref{dsmisner}, one can find that only the difference is the power of $\lambda$ in the metric components. This is the reason why the structure of two spacetimes are very similar to each other. We note that similar Misner type of singularity was also known in the compactified ${\it open}$ de Sitter space~\cite{Ishibashi:1996ps}.

In some examples, quasi regular singularities are associated with the fixed points of the identification (or the orbifold action) when such singular geometries come from the quotient of original spacetime. To see an interesting difference from the Misner space, let us study the fixed points of the torus identification in our cases.
Here, the torus identification is expressed as the integral of infinitesimal translations in $x$ direction in the flat chart coordinate, it is sufficient to check fixed points of the generating vector $\partial_{x}$.
In the coordinate $\{\tau, \chi \}$, the vector $\partial_x$ becomes
\begin{eqnarray}
\partial _x &=& \frac{\partial \tau}{\partial x} \partial_{\tau} +\frac{\partial \chi}{\partial x} \partial_{\chi} \notag \\
&=& H \sin \chi \cos \tau \partial_{\tau} + H (1 + \cos \chi \sin \tau) \partial_{\chi}.
\end{eqnarray}
At fixed points, we get 
\begin{equation}
\sin \chi \cos \tau = 0 ,\qquad 1 + \cos \chi \sin \tau = 0.
\end{equation}
The latter condition imposes $(\chi, \tau) = (0, -\frac{\pi}{2})$ or $(\chi, \tau) = (\pi, \frac{\pi}{2})$.
The former condition is also satisfied at these points.
They are located on the asymptotic boundaries and there are no fixed point in the bulk of de Sitter.
This is in contrast to Misner spacetime.
We can obtain Misner spacetime by taking the orbifold of Minkowski spacetime and they have a fixed point in the bulk of Minkowski spacetime.

\section{Summary and Discussions}
In the present paper, we discussed the spacetime structure of extended torus de Sitter universe. 
First we demonstrated that the torus de Sitter universe has past incomplete null geodesics, which is consistent with the result in Refs.~\cite{Galloway:2004bk, Galloway:2017mts}. Then we presented an extension of torus de Sitter universe where an incomplete geodesic in the original region becomes past complete. We found, however, that the resultant spacetime has a closed causal curve. In addition we found that another incomplete curve in original region is still incomplete even after the extension. We proved that this incomplete curve is inextendible, i.e. extended torus de Sitter Universe has a singularity, though it is very mild in the sense that there is no divergence of curvature there. we would like to note that, in the course of analysis, we only assumed the periodicity of only one direction $x$. Thus our result can be applied not only for torus Universe but also cylindrical Universe, where one of other direction remains uncompactified .

We should stress that our results do not prohibit the presence of the inflationary epoch in the model of torus universe because all ill behavior appear in the limit $t \rightarrow - \infty$. Rather, our result suggests that there must be non-inflationary phase before the inflationary one, if one believes that no singularity exists in the real Universe. For example, it is pointed out that the Casimir energy behaves like radiation in the torus universe model and hence there would be the radiation dominant era before inflation \cite{Fornal:2011tw,Fornal:2012kx}, though in this case we will hit the big bang singularity.
Quantum creation of compact universe \cite{Zeldovich:1984vk,Coule:1999wg, Linde:2004nz} is also a possible scenario to avoid our mild singularity because semi-classical description is no longer valid near $a = 0$.
 
Incompleteness of our torus universe is also interesting in the perspective of quantum gravity in de Sitter, especially in the context of holography \cite{Witten:2001kn,Strominger:2001pn}.
If such quantum gravity
 exists, we will be able to consider quantum gravity in de Sitter that is asymptotically torus \cite{Castro:2011xb}.
In the semiclassical limit, a wave function is
 expanded as sum over classical solutions and fluctuations on them.
However it is highly nontrivial what kind of classical
 geometries should be included in the semiclassical expansion. 
In Euclidean quantum gravity in anti de Sitter, there are arguments that the singular geometries should be included \cite{Maloney:2016gsg}.
Therefore, 
it is natural to expect that our singular torus de Sitter universe must be included as well in the case of de Sitter quantum gravity with compactified boundary.
Even so,
it is not manifest which region of torus universe is included: as we explained in the footnote \ref{FN2}, one can construct complete torus de Sitter Universe if one allows non-Hausdorff spacetime. 
It will be worth studying these problems to understand quantum gravity.

\begin{acknowledgments}
T.N. is supported by JSPS fellowships and the Simons Foundation through the It From Qubit collaboration.
D.Y. is supported by the JSPS Postdoctoral Fellowships for Research Abroad.
\end{acknowledgments}
\appendix
\section{Coordinates in de Sitter space}
$d$ dimensional de Sitter space $dS_d$ is realized as a hyperboloid in Minkowski space $\mathbb{R}^{1,d}$.
The metric in $\mathbb{R}^{1,d}$ is given by the usual one:
\begin{equation}
ds_{embed}^2 = -dX_0^2 + dX_1^2 + \cdots + dX_d ^2.
\end{equation}
de Sitter space with radius $1/H$ is the hypersurface defined by the following equation:
\begin{align}
-X_0^2 + X_1 ^2 + \cdots +X_d^2 = \frac{1}{H^2}.
\end{align}
The coordinate system that covers entire de Sitter space is given by
\begin{align}
X_0 = \frac{1}{H} \tan \tau, \qquad X_i = \frac{1}{H\cos \tau} \Omega_i , \label{eq:globalcoord}
\end{align}
where $i = 1 ,\cdots, d$ and $\Omega_i$ lives on a sphere with the unit radius i.e. $\vec{\Omega}^2 = 1$.
The metric is induced from the embedding space $\mathbb{R}^{1,d}$, and given by
\begin{align}
ds^2_{global} =  \frac{1}{H^2 \cos \tau^2} (- d \tau^2 + d \Omega_{d-1} ^2).
\end{align}
On the other hand, the flat chart coordinate is given by the following embedding:
\begin{align}
X_0 &= \frac{1}{H} \sinh (Ht) + \frac{Hr^2}{2} e^{H t} , \notag \\
X_1 &=  \frac{1}{H} \cosh (Ht) - \frac{Hr^2}{2} e^{H t}, \notag \\
X_j &= e^{Ht} x_j, \label{eq:flatcoord}
\end{align}
where $j = 2, \cdots, d$ and $\vec{x}^2 = r^2$. 
This coordinate only covers a half of whole de Sitter space that satisfies $X_0 + X_1 > 0$.
The metric is also induced from the embedding space $\mathbb{R}^{1,d}$, and given by
\begin{equation}
ds^2_{flat} = - dt^2 + e^{2 Ht }d\vec{x}^2.
\end{equation}
This is nothing but the flat chart metric.
By comparing (\ref{eq:globalcoord}) and (\ref{eq:flatcoord}), we can find the relation between the flat chart coordinate and that of global chart.
In $d = 2$ case, after putting $\Omega_1 = \cos \chi$ and $\Omega_2 = \sin \chi$, we obtain the relation (\ref{eq:ds2coordchange}) in the bulk.

\bibliography{ref}
\end{document}